\documentclass[pra,aps, twocolumn, showpacs,groupedaddress,10pt,nofootinbib]{revtex4-1} 
\usepackage[T1]{fontenc}
\usepackage[utf8]{inputenc}
\usepackage{soul}
\usepackage{enumitem}
\usepackage{graphicx}
\usepackage{dcolumn}
\usepackage{bm}
\usepackage[draft]{todonotes}
\usepackage{xcolor}
\usepackage{comment}
\usepackage[normalem]{ulem}
\usepackage{appendix}

\begin{document}
\renewcommand{\appendixname}{APPENDIX}
\title{Realization of a stroboscopic optical lattice for cold atoms with subwavelength spacing}

\author{T-C. Tsui}
\thanks{These two authors contributed equally}
\affiliation{Joint Quantum Institute, National Institute of Standards and Technology and the University of Maryland, College Park, Maryland 20742 USA}

\author{Y. Wang\normalfont\textsuperscript{*,}}
\email{Corresponding author. wang.yang.phy@gmail.com}
\affiliation{Joint Quantum Institute, National Institute of Standards and Technology and the University of Maryland, College Park, Maryland 20742 USA}

\author{S. Subhankar}
\affiliation{Joint Quantum Institute, National Institute of Standards and Technology and the University of Maryland, College Park, Maryland 20742 USA}

\author{J. V. Porto}
\affiliation{Joint Quantum Institute, National Institute of Standards and Technology and the University of Maryland, College Park, Maryland 20742 USA}

\author{S. L. Rolston}
\affiliation{Joint Quantum Institute, National Institute of Standards and Technology and the University of Maryland, College Park, Maryland 20742 USA}

\date{\today}
\begin{abstract}
Optical lattices are typically created via the ac-Stark shift, which are limited by diffraction to periodicities  $\ge\lambda/2$, where $\lambda$ is the wavelength of light used to create them.  Lattices with smaller periodicities may be useful for many-body physics with cold atoms and can be generated by stroboscopic application of a phase-shifted lattice with subwavelength features. Here we demonstrate a $\lambda/4$-spaced lattice by stroboscopically applying optical Kronig-Penney(KP)-like potentials which are generated using spatially dependent dark states.  We directly probe the periodicity of the $\lambda/4$-spaced lattice by measuring the average probability density of the atoms loaded into the ground band of the lattice. We measure lifetimes of atoms in this lattice and discuss the mechanisms that limit the applicability of this stroboscopic approach.  
\end{abstract}


\maketitle
\section{Introduction}
Ultracold atoms trapped in periodic optical potentials provide wide-ranging opportunities to study many-body physics in highly controllable systems \cite{Maciej2007,Bloch2008}. In all cases, the characteristic single-particle energy scale is set by the recoil energy, $E_R = h^2 /(8md^2)$, where $m$ is the mass of the atom and $d$ is the spatial period of the lattice. Although temperatures in such systems can be quite low, it is still challenging to reach temperatures well below the relevant  many-body physics energy scales, which can be exceedingly small. Increasing the recoil energy can potentially increase both single-particle and many-body energy scales through tighter confinement, which may aid in creating systems well into the regime where many-body ground state physics is observable.  An inherent obstacle to smaller lattice spacing is the optical diffraction limit, which prevents lattice periodicities below $d = \lambda/2$, where $\lambda$ is the wavelength of the light forming the lattice.  
Several approaches to move beyond the diffraction limit have been proposed and some realized based on multiphoton effects \cite{Dubetsky2002,Ritt2006,Anderson2019}, rf-dressed adiabatic potentials \cite{Yi2008,Lundblad2008,Lundblad2014}, and 
trapping in near-field guided modes with nanophotonic
systems \cite{Gullans2012,Thompson2013,Romero-Isart2013,Gonzalez2015}.  \\   

\begin{figure}[t]
\includegraphics[height=2.8in]{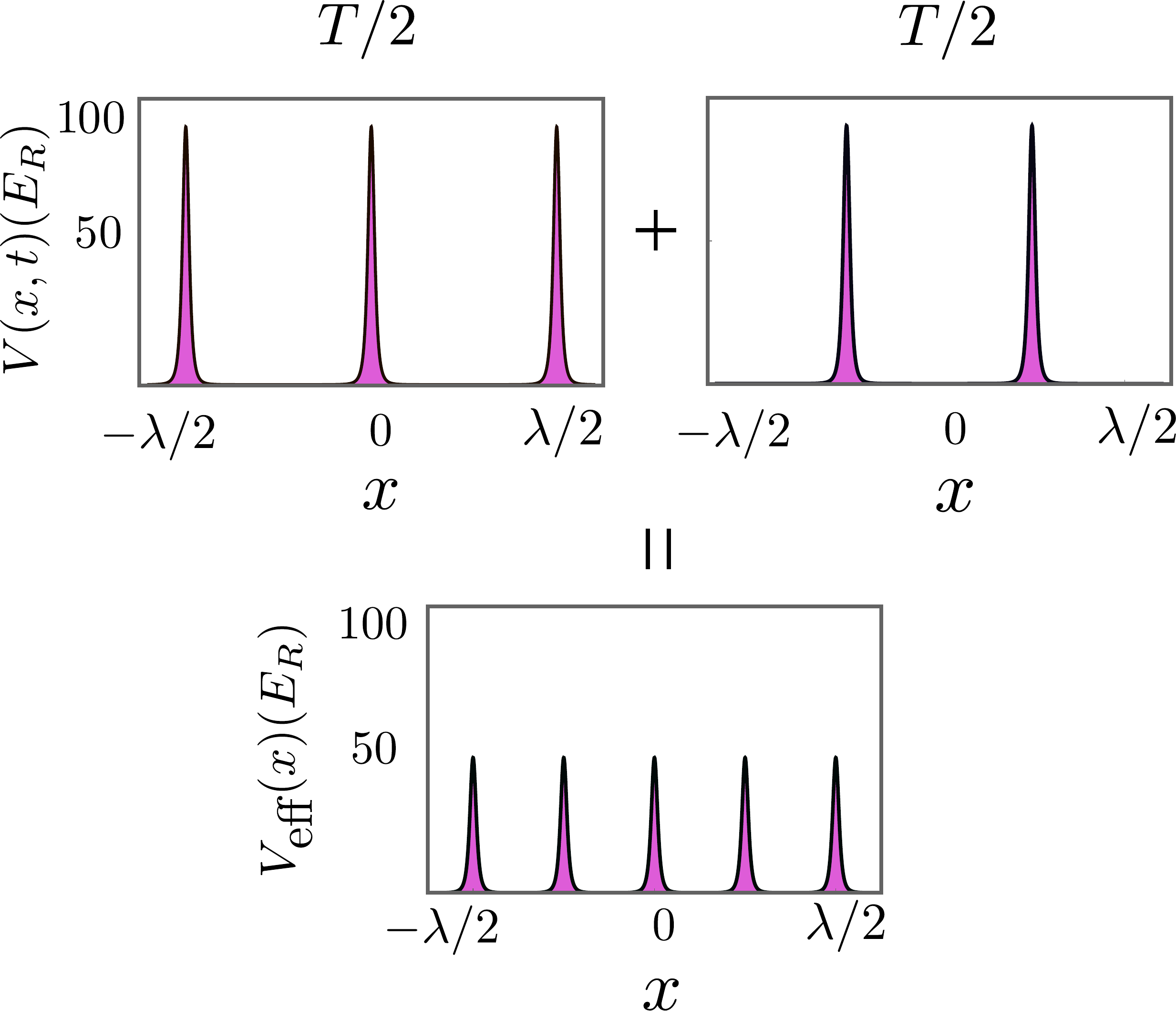}
\caption{The stroboscopic approach to create a time-averaged effective potential with a lattice spacing of $\lambda/4$ by dynamically pulsing KP potentials with $\lambda/2$ spacing.}
\label{fig:schematic} 
\end{figure}
Here we report the realization of a recently proposed Floquet-based approach~\cite{Nascimbene2015,Subhankar2019b,Lacki2019}  to create small-period lattices, specifically $\lambda/4$-spaced lattices, by time-averaging a modulated lattice potential that has subwavelength features. We load atoms into the ground band of this time-dependent lattice and measure their average probability density $|\psi_{\text{avg}}(x)|^2$ with nanoscale resolution \cite{Subhankar2019,McDonald2019,Tonyushkin2006}, to confirm the subwavelength nature of the lattice. We study the lifetime of atoms in the lattices over a range of modulation (Floquet) frequencies $\omega_F=2\pi/T$, where $T$ is the period of a complete cycle, to determine the frequency range over which the time-averaged approach works. 

Creating an effective time-averaged potential requires that the time-dependence of the lattice be motionally diabatic \cite{Eckardt2017,Rahav2003,Marin2015}, namely that $T$ is much smaller than the motional time scale of the atoms.
Time-averaging a dynamically applied lattice potential cannot create an effective potential landscape with higher spatial Fourier components than the underlying progenitor lattice. This implies that in order to create landscapes with subwavelength periodicity, one must time-average a potential that itself has subwavelength features~\cite{Nascimbene2015}. In this work, we make use of the  Kronig-Penney(KP)-like potential to generate the desired potential landscapes~\cite{Subhankar2019b,Lacki2019}. Such a KP potential is implemented via the dark state associated with a three-level $\Lambda$-system \cite{Lacki2016,Jendrzejewski2016,Wang2018}. The spin adiabaticity required to maintain the dark state during the stroboscopic cycle imposes additional constraints, as discussed below.
\begin{figure}[t]
\includegraphics[width=8.6cm]{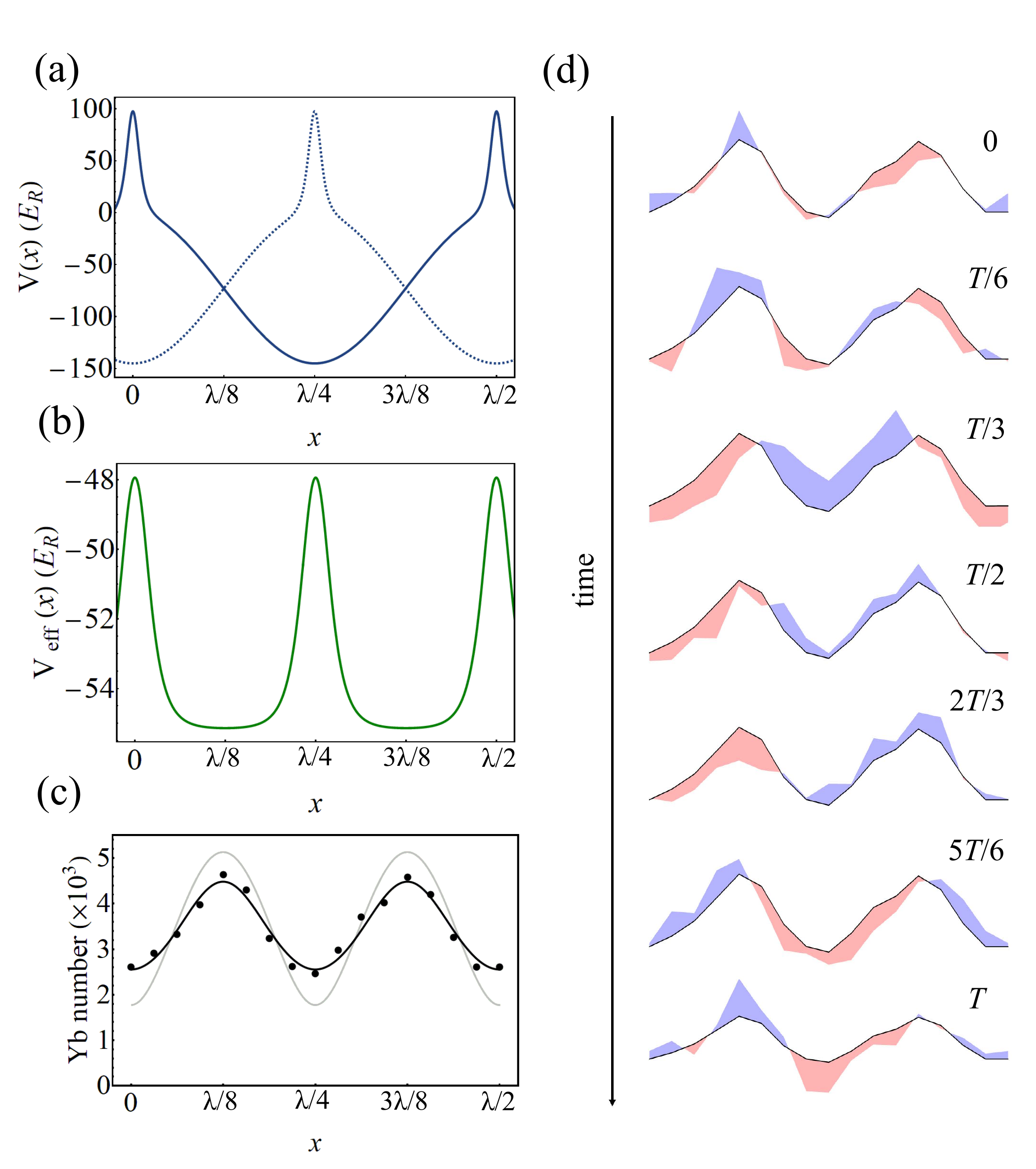}
\caption{\label{fig:wavefunction}(a)  The stroboscopically applied potential, shown here for $\Omega_{c0}=500\Gamma$ and $\Omega_{p}=50\Gamma$, is composed of KP barriers on top of a sinusoidal potential. The dotted line represents the potential shifted by $\lambda/4$. (b) The time-averaged effective potential $V_{\text{eff}}(x)$. (c) The black points are the measured $|\psi_{\text{avg}}(x)|^2$ of atoms in $V_{\text{eff}}(x)$. Number fluctuations between realizations result in number uncertainties of 5\%. The black line is the calculation based on independently measured lattice parameters. The grey line is the calculated $|\psi_{\text{avg}}(x)|^2$ in the lattice before the relaxation during the measurement. (d) The micromotion dynamics at different time within a Floquet period. The blue(red)-shaded areas  represent regions in which $|\psi(x,t)|^2$ is higher(lower) than $|\psi_{\text{avg}}(x)|^2$, which is shown as a solid black line. }
\end{figure}

There are multiple ways to implement time-averaging with a KP lattice \cite{Lacki2019,Subhankar2019b}. The particular approach that we adopt, optimized for our experimental conditions, is shown in Fig.\ref{fig:schematic}. Periodic potentials with $\lambda/2$ spacing but subwavelength structure are stroboscopically applied to the atoms to create the desired potential landscape. Specifically,
atoms are subjected to a KP potential for half of the Floquet cycle $T/2$;
the potential is then ramped down to zero and its position is shifted by half of the lattice spacing $\lambda/4$;
the shifted potential is ramped on again and held for another half cycle, before being ramped off and its position is restored. 

Two factors must be considered to ensure that time-averaging is an effective description of the system. First, motional diabaticity sets a lower bound on the Floquet frequency $\omega_F$, beyond which the band structure becomes unstable and severe heating limits the lifetime. Second, the dark-state nature of the KP lattice sets an upper bound to $\omega_F$. As the  KP potential is a scalar gauge potential arising from a spatially varying dark state~\cite{Wang2018,Lacki2016,Jendrzejewski2016}, switching on and off such a potential requires atoms to adiabatically follow the spatio-temporal dark state at all times. We ensure this adiabatic following by carefully designing the pulse shapes of our light fields (Appendix~C), implementing stimulated Raman adiabatic passage (STIRAP) \cite{Vitanov2017}. Losses occur at high $\omega_F$, as the atom's dark-state spin composition fails to adiabatically follow the rapid changes in the light fields. In the following sections, we show that a frequency window that simultaneously satisfies both requirements exists
 and that there are momentum-dependent loss channels arising from the Floquet-induced coupling with higher excited bands for particular momenta.  
\section{Experiment}
We work with fermionic $^{171}$Yb atoms that have a well isolated $\Lambda$-system (Appendix~A), consisting of two ground states $|g_1\rangle$, $|g_2\rangle$ 
and an excited state $|e\rangle$ coupled by laser light with $\lambda=556$ nm. We use the methods outlined in Refs.~\cite{Wang2018,Vaidya2015,Pisenti2016, Subhankar2019c,Appel2009} to generate and optically control this well isolated $\Lambda$-system. A  control field  $\Omega_c(x,t)=\Omega_{c1}e^{ikx}+\Omega_{c2}(t)e^{-i(kx+\phi(t))}$, where $k=2\pi/\lambda$ and $\phi(t)$ is the relative phase difference between the two fields, which couples $|g_2\rangle$ and $|e\rangle$, is comprised of two counterpropagating lattice beams. The maximum value of $\Omega_{c2}(t)$ is constrained to be equal to $\Omega_{c1}=\Omega_{c0}/2$, in which case it gives rise to a standing wave $\Omega_{c0}\, e^{-i\phi(t)/2}\cos {(kx +\phi (t)/2)}$. We control the strength and the position of the KP potential using $\Omega_{c2}(t)$ and $\phi(t)$ (Appendix~C). A homogeneous probe field $\Omega_p e^{iky}$, coupling $|g_1\rangle$ and $|e\rangle$, travels perpendicular to the control beams. The resulting spatially dependent dark state gives rise to a KP lattice of narrow subwavlength barriers~\cite{Lacki2016,Jendrzejewski2016,Wang2018}, plus an additional sinusoidal potential due to the light shifts caused by states outside the three-level system (Appendix~A) as shown in Fig.~\ref{fig:wavefunction}(a).

Stroboscopically applying the lattice with different strengths and positions requires accurate and high bandwidth control of the amplitude and phase of the lasers coupling the three states, which we implement using dynamic control over the rf fields driving acousto-optic modulators (AOMs)~\cite{Subhankar2019}.  We note that the spin adiabaticity condition depends significantly on the pulse shape~\cite{Subhankar2019b} in addition to the Floquet frequency, and control of the pulse shape within a Floquet period is critical~\cite{Subhankar2019}. We use arbitrary waveform generators that can control the rf amplitude and phase with a resolution of 8 ns and 4 ns respectively. However, we are limited by the bandwidth of the AOMs, which we measure to be 50 ns. This is a factor of 8 times smaller than the smallest half-period of 400 ns that we have used in this study. 

For typical experimental values of $\Omega_{c0}=500$~$\Gamma$ and $\Omega_p=50$~$\Gamma$, where $\Gamma=2\pi\times182$~kHz is the inverse lifetime of $|e\rangle$, the KP barrier has a minimum width of 0.02~$\lambda$ and a maximum height $\approx 100 E_R$, where $E_R/h=h/(2m_{\text{Yb}} \lambda^2)=3.7$~kHz, $m_{\text{Yb}}$ is the mass of a $^{171}$Yb atom, and the sinusoidal potential has a depth $\approx 145 E_R$, Fig.~\ref{fig:wavefunction}(a). Time-averaging this lattice applied at two positions results in an effective potential $V_{\text{eff}}(x)$ shown in Fig.~\ref{fig:wavefunction}(b), which includes the effect of the pulse shapes, with an effective barrier height $\approx7 E_R$. (The sinusoidal component of the potential averages to a spatially invariant offset.)

\section{Measurement}
We apply this lattice to $\approx 2\times10^5$ Yb atoms at an initial temperature of 0.3 $\mu$K that have been optically pumped into $|g_1\rangle$. 
To load the atoms into the ground band of $V_{\text{eff}}(x)$, we adiabatically increase the depth of the stroboscopically applied lattices in 200 $\mu s$ (typically $\sim$80 Floquet cycles) described in details in Appendix~B. After the loading stage, we  measure the ensemble-averaged probability density $|\psi(x,t)|^2$ of atoms in the ground band of $V_{\text{eff}}(x)$ using a nanoresolution miscroscopy technique~\cite{Subhankar2019} with FWHM resolution of 25~nm. We also measure the momentum distribution of the atoms via absorption imaging after time-of-flight (TOF).

\subsection{Probing wavefunction density in the stroboscopic lattice}
Figure \ref{fig:wavefunction}(c) shows $|\psi(x,t)|^2$ averaged over a Floquet period $T=2.4\,\mu s$ ($\omega_F=2\pi \times 410$ kHz) for atoms in $V_{\text{eff}}(x)$ with a $\lambda/4 $ lattice spacing, and Fig.~\ref{fig:wavefunction}(d) shows $|\psi(x,t)|^2$ at different times within a Floquet cycle. By averaging the data over a Floquet period, we eliminate the effect of micromotion and obtain the averaged wavefunction density $|\psi_{\text{avg}}(x)|^2$ (dotted trace in Fig.\ref{fig:wavefunction}(c)) in the ground band of the effective potential. The black curve represents the ground-band probability density calculated from the time-averaged potential including the quasimomentum averaging, the effect of finite resolution of the microscope, and the relaxation of the wavefunction during the measurement. The good agreement between the data and calculation shows that time-averaging is a good description of the effective potential. The calculated wavefunction in the lattice before the relaxation during the measurement is plotted in grey. We resolve the micromotion in real space within a Floquet period by comparing $|\psi(x,t)|^2$ with $|\psi_{\text{avg}}(x)|^2$ (Fig.\ref{fig:wavefunction}(d)). The blue (red)-shaded areas represents regions in which $|\psi(x,t)|^2$ is higher (lower) than $|\psi_{\text{avg}}(x)|^2$. We observe that micromotion has the same time-periodicity as the Floquet drive, as expected.

\begin{figure}
\includegraphics[width=8.6cm]{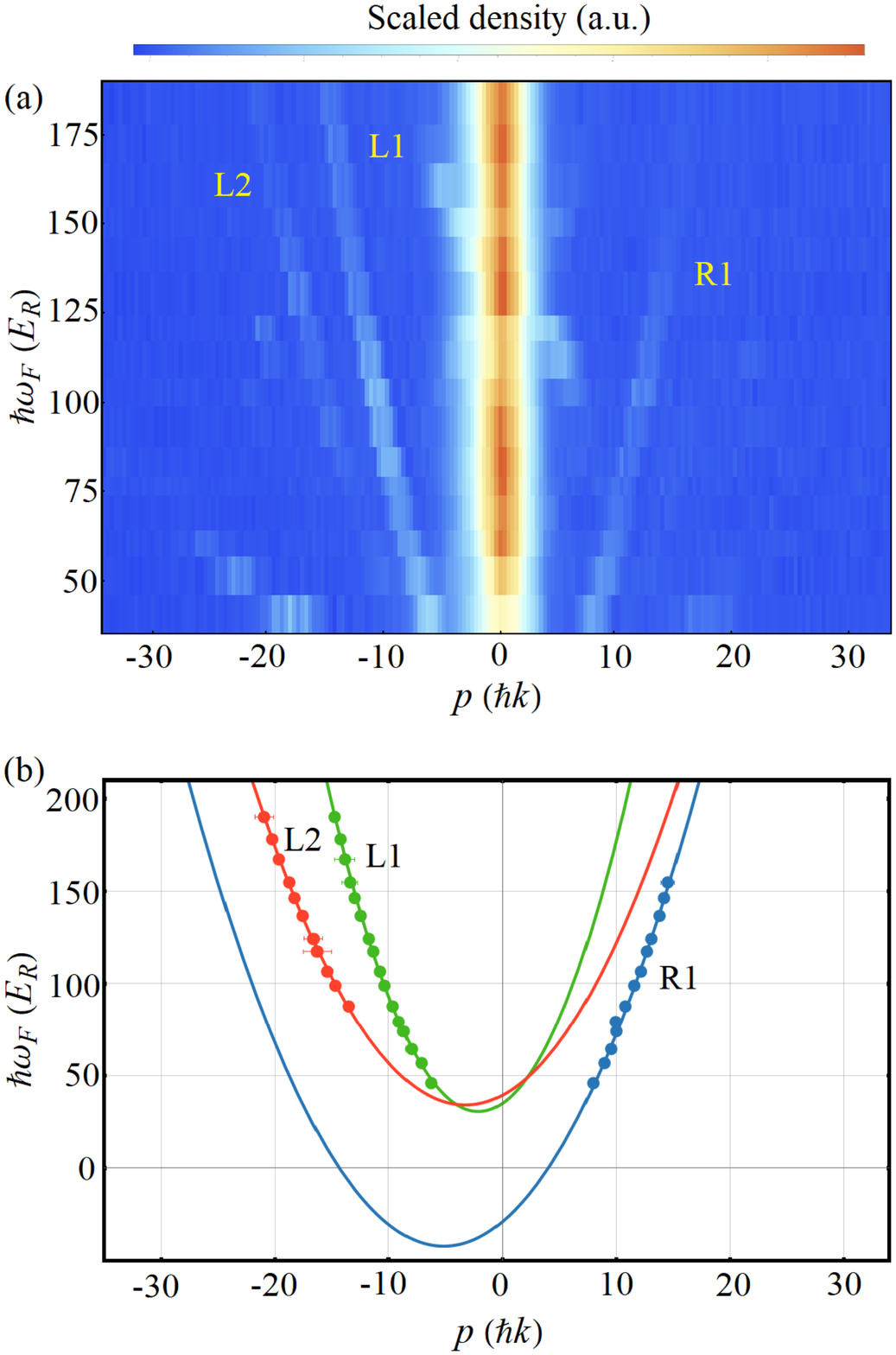}
\caption{\label{fig:TOF}(a) Integrated TOF column density at different Floquet frequencies  $\omega_F$. The atomic populations at high momenta indicate the presence of avoided crossings. The widths of the populations at avoided-crossings are primarily due to the physical dimensions of the atomic cloud. (b) The Floquet frequency $\omega_F$ is plotted versus the center momentum of the populations in (a) determined using Gaussian fits. Different series of avoided crossing are labeled and colored (L1: green, L2: red, R1: blue) and their fitted quadratic functions are drawn in solid lines respectively. The error bars are 1 standard deviation of the Gaussian fits. }
\end{figure}
\subsection{Momentum-dependent loss channels}
A characteristic feature of a Bloch-Floquet bandstructure is the existence of avoided crossings at particular lattice momenta arising from coupling with high-lying states \cite{Holthaus_2015}, which for large Floquet frequency are approximately plane waves with high momenta. We measure the momentum distribution of the atoms in $V_{\text{eff}}(x)$ at different $\omega_F$ by taking an absorption image after ramping down the lattice in 100 $\mu s$ followed by a TOF of 3~ms. The atomic populations at high momenta in Fig.\ref{fig:TOF}(a) indicate the mixing of low momentum and high momentum states due to the presence of avoided crossings in our system. We use a Gaussian fit to determine the center momentum of the populations with respect to the ground band. The Floquet frequency $\omega_F$ is plotted against the center momentum (Fig.\ref{fig:TOF}(b)) for the three most prominent peaks (L1: green, L2: red, R1: blue).  To first order, the avoided crossings can be understood as arising from the crossing of Floquet dressed high-lying bands, which are shifted in energy by integral multiples of $\omega_F$, and the low-lying occupied bands of $V_{\text{eff}}(x)$, which are relatively flat. To determine the integral multiple of $\omega_F$ for the band coupling, we fit the peak positions with a quadratic function $\hbar \omega_F=(p-p_0)^2/N +\hbar \omega_0$, where $p$ is the momentum, $N$ is an integer, $p_0$ and $\omega_0$ are fitting parameters, and the momentum and energy are in units of $\hbar k$ and $E_R$. For the L1 series, a good agreement with the data is found for $N=1$, indicating this series is due to coupling between bands with an energy difference of $\hbar\omega_F$. For the L2 and R1 series, $N=2$ gives the best fit, indicating second order coupling between bands that differ in energy by 2$\hbar\omega_F$. (The other visible peaks do not extend over a sufficient range to accurately determine their curvatures.)  The fraction of atoms in the high momentum states decreases at higher Floquet frequency, suggesting weaker coupling to higher bands. The asymmetry in the avoided crossings with respect to $p=0$ is due to the fact that we are driving just the $\Omega_{c2}$ control beam, which gives rise to a vector gauge potential~\cite{Subhankar2019b}.

The $^{171}$Yb atoms are nearly non-interacting ($s$-wave scattering length is $-3a_0$, where $a_0$ is the Bohr radius), so they are not likely to thermalize during the short loading and unloading sequence.  However, the observed low-momentum component of the TOF distribution is consistent with the width of the ground band Brillouin zone for the $\lambda/4$-spaced stroboscopic lattice, which is twice as large as the ground band width of the progenitor $\lambda/2$ lattice. Given that the Fermi momentum at our density is of order the recoil momentum of the progenitor lattice, the filled ground band in the $\lambda/4$ lattice indicates that the effective temperature is higher than the ground band width but not a significant fraction of the band spacing.

\begin{figure}[t]
\centering
\includegraphics[width= 8.6cm]{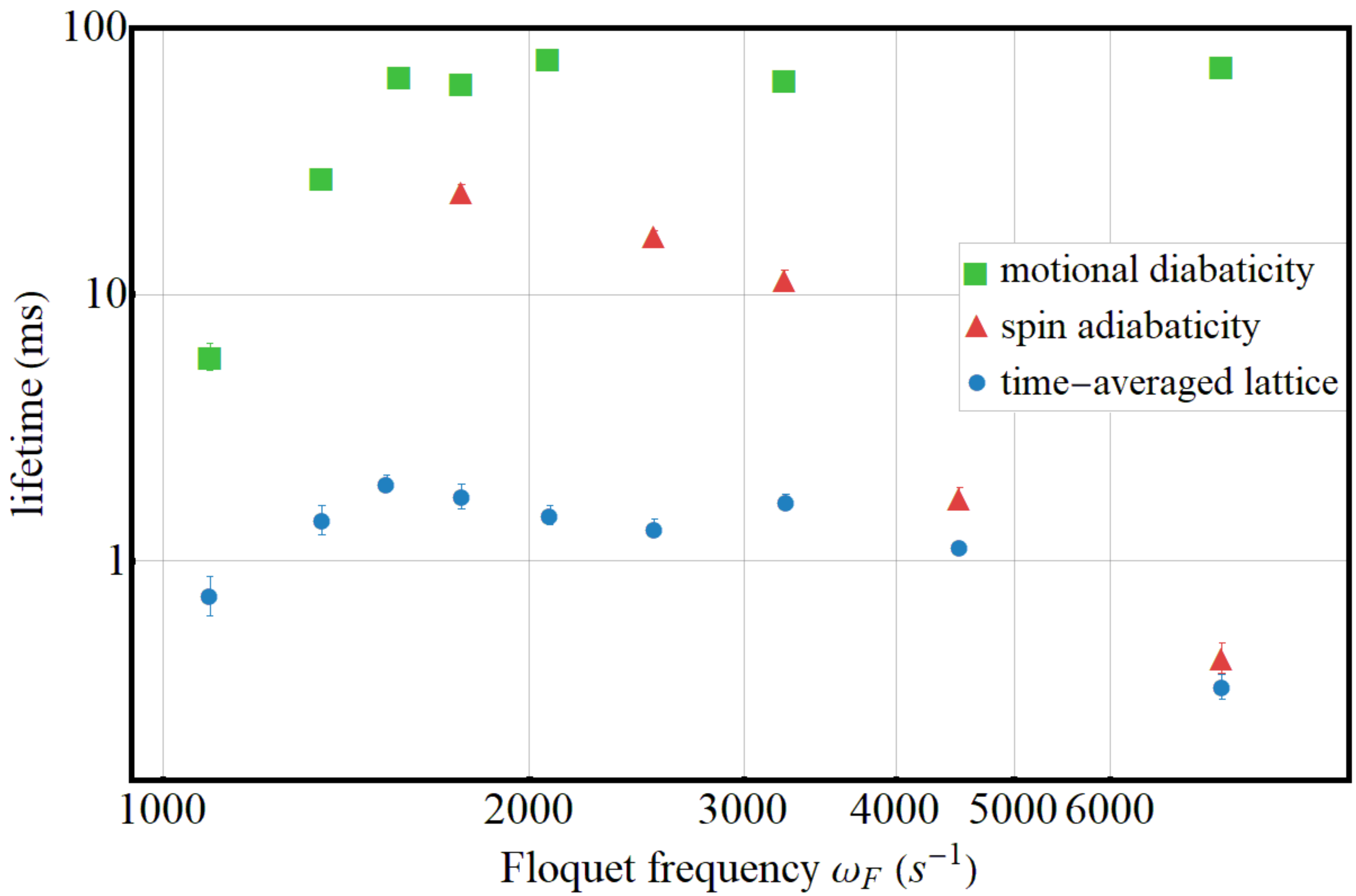}
\caption{\label{fig:lifetime} Lifetimes of atoms at different $\omega_F$ under different Rabi frequency configurations. Green squares: $\Omega_{c0}=500\Gamma$ and $\Omega_p=0$, where the spin degree of freedom is decoupled and the loss is due solely to failure of motional diabaticity at low $\omega_F$. Red triangles: $\Omega_{c1}=0$, $\Omega_{c2}=250\Gamma$ and $\Omega_p=80\Gamma$, where the spatial potential is homogeneous and the loss is due solely to the failure of spin adiabaticity at high $\omega_F$. Blue circles: $\Omega_{c0}=500\Gamma$ and $\Omega_p=80\Gamma$, where we show the lifetimes of atoms in the $\lambda/4$-spaced lattice, $V_{\text{eff}}(x)$. The error bars are 1 standard deviation of the exponential fits.}
\end{figure}

\subsection{Lifetime study}
In order to determine the range of usable Floquet frequencies for the stroboscopic scheme, we study the lifetime at different $\omega_F$ under different Rabi frequency configurations as shown in Fig.~\ref{fig:lifetime}. We determine the lower bound on $\omega_F$ by studying the motional diabaticity of atoms in just a stroboscopically applied ac-Stark-shift lattice. This is done by setting $\Omega_p=0$, which decouples the spin degree of freedom from the dynamics with $\Omega_{c1}=250\Gamma$, while $\Omega_{c2}(t)$ is pulsed to a maximum value of $250\Gamma$(Appendix~C). At low $\omega_F$, the atoms are affected by the turning on and off, and phase shifting of the sinusoidal ac-Stark-shift potential, which causes heating and loss (green squares in Fig.\ref{fig:lifetime}). We determine the upper bound on $\omega_F$ by studying the reduction in the fidelity of STIRAP as a function of $\omega_F$ for a spatially homogeneous dark state. This is done by setting $\Omega_{c1}=0$, $\Omega_p=80\Gamma$, while $\Omega_{c2}(t)$ is pulsed to a maximum value of $250\Gamma$. The reduction in STIRAP fidelity manifests as heating and loss due to the decreasing spin adiabaticity at larger $\omega_F$. Most importantly, we also measure the frequency dependent lifetime of atoms loaded into $V_{\text{eff}}(x)$ for different $\omega_F$ (blue circles in Fig.\ref{fig:lifetime}). The reduction in spin adiabaticity accounts for the decrease in lifetime of atoms in $V_{\text{eff}}(x)$ at high $\omega_F$.

The short lifetimes in the stroboscopically applied KP lattices are expected due to a few factors. First, couplings to the spatially and temporally dependent bright states reduce lifetimes in subwavelength-spaced lattices even for a perfect three-level system, through couplings with higher Floquet bands (as shown in Fig.~\ref{fig:TOF}) and off-resonant couplings with bright states~\cite{Subhankar2019b}. In principle, these couplings can be reduced by using larger Rabi frequencies. However, lifetimes are also limited by the breakdown of the three-level approximation at large Rabi frequencies due to admixing of states outside the three-level system (Appendix~A). This manifests as a dynamically varying and spatially dependent two-photon detuning (arising from $\Omega_c(x,t)$), which reduces the fidelity of STIRAP~\cite{Vitanov2017}. This competing requirement prevents us from benefiting from larger Rabi frequencies.

\section{Conclusion}
In conclusion, we demonstrate the creation of a time-averaged $\lambda/4$-spaced lattice using a recently proposed stroboscopic technique~\cite{Nascimbene2015} based on dynamically modulated dark states in a three-level system~\cite{Subhankar2019b,Lacki2019}. The subwavelength structure of the lattice is confirmed by measuring the probability density of the atoms averaged over the ground band of the lattice.  We measure the loss rate of atoms in the lattice and observe high momentum excitation due to Floquet-induced coupling to higher bands. We measure the lifetime of the atoms in the $\lambda/4$-spaced lattice to be 2~ms,  which is not long enough compared to the tunneling time to allow for many-body studies in the current realization. 

Further improvement of the $\lambda/4$-spaced lattice would require compensation of the two-photon detuning\st{s} or the identification of other atomic systems with a more favorable (isolated) three-level system~\cite{Bienias2018}. 
The lattice demonstrated here is limited by the off-resonant coupling to $|(6s6p)^3P_1,F$=$3/2,m_F$=$-3/2\rangle$, which is only detuned by the hyperfine splitting from the three-level system being used. Better candidates may make use of isolated {\em electronic} levels, which are detuned by much larger optical separations.
For example, in $^{174}$Yb, the $(6s6p)^3P_0$ state and one of the states in the $(6s6p)^3P_2$ level could be used as the ground states, while one of the $(6s7s)^3S_1$ states could be used as the excited state, with appropriate choice of polarization to select the three states. In a more isolated three-level system the main limitation would be the available laser power needed to meet the Rabi frequency requirements. In addition to longer lifetimes, higher Rabi frequencies would allow for lattices with smaller spacings~\cite{Subhankar2019b}. Our work can be extended to 2D and additional dynamic control over the two-photon detuning---which makes subwavelength traps possible~\cite{Bienias2018}---allows for construction of arbitrary time-averaged potential landscapes not limited by diffraction.
\section*{ACKNOWLEDGMENTS}
We acknowledge support from NSF PFC at JQI (Grant No.
PHY1430094) and ONR (Grant No. N000141712411).

\section*{APPENDIX}
\subsection{$^{171}$Yb ATOM LEVEL STRUCTURE}
\label{LevelStructure}
\begin{figure}[h]
	\includegraphics[width=8.6cm]{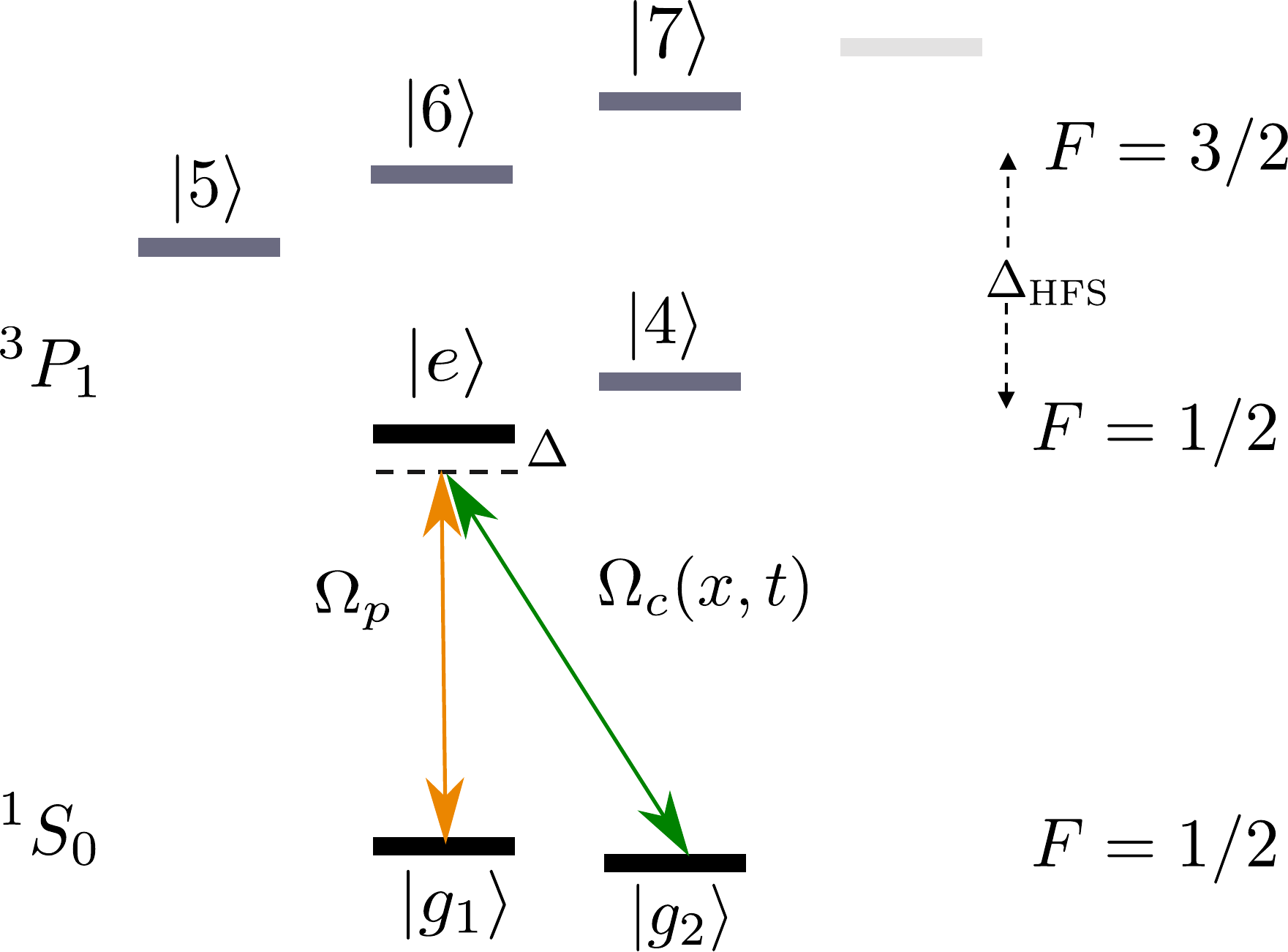}
	\label{fig:LevelDiagram}
	\caption{Level structure of the $^1S_0$ and $^3P_1$ manifolds in $^{171}$Yb: $\Delta$ is the single photon detuning, and $\Delta_{\text{HFS}}\approx 6$ GHz is the $^3P_1$ hyperfine splitting.
	}
\end{figure}
\begin{figure*}[]
	\centering
	\includegraphics[width=17.8cm]{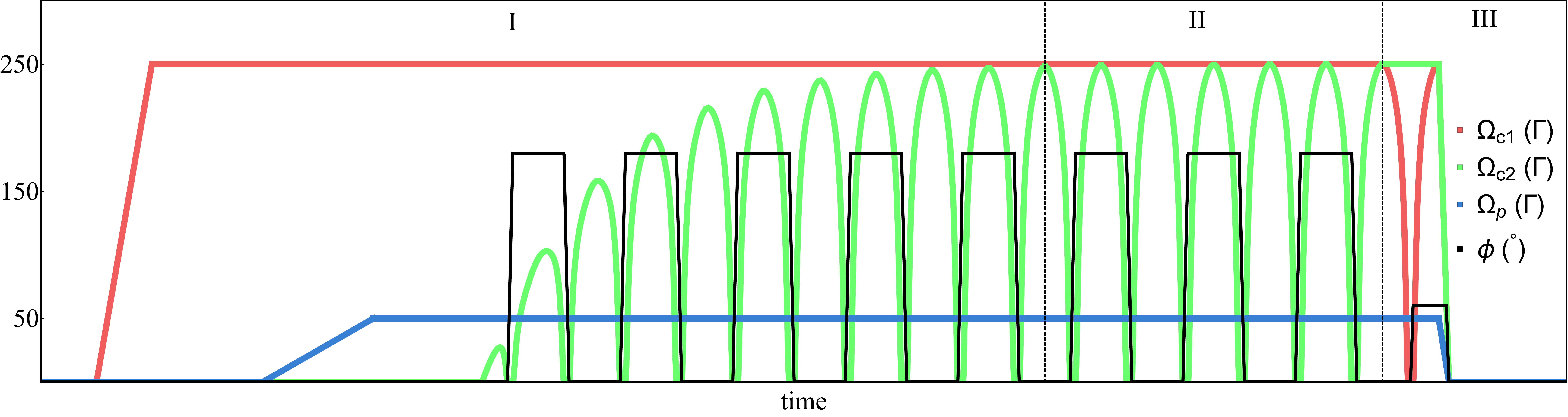}
	\label{fig:Sequence} 
	\caption{ Rabi frequencies of different light fields and the relative phase $\phi$ between $\Omega_{c1}$ and $\Omega_{c2}$ during three stages. The Floquet period is not shown to scale, the minimum number of Floquet cycles during the ramp-on of $\Omega_{c2}$ is 40.}
\end{figure*}
Fig. 5 shows the level structure of the $^1S_0$ and $^3P_1$ manifolds in $^{171}\text{Yb}$. The three hyperfine states $|g_1\rangle$, $|g_2\rangle$, and $|e\rangle$ constitute the $\Lambda$-system. We use a magnetic field of 36~mT to yield a frequency separation of 1 GHz between $|e\rangle$ and $|4\rangle$. The hyperfine splitting is $\Delta_{\text{HFS}}\approx6$~GHz. 

The ac-Stark shifts on the ground states $|g_1\rangle$ and $|g_2\rangle$ arise due to off-resonant couplings to states outside the $\Lambda$-system. The $\Omega_c(x,t)$ light field off-resonantly couples $|g_1\rangle$ with $|5\rangle$, and $|g_2\rangle$ with $|6\rangle$. The $\Omega_p$ light field off-resonantly couples $|g_2\rangle$ with $|4\rangle$, $|g_2\rangle$ with $|7\rangle$, and $|g_1\rangle$ with $|6\rangle$. The spatio-temporally dependent ac-Stark shifts due to $\Omega_c(x,t)$ give rise to the dynamic sinusoidal potential mentioned in the main text.
\subsection{EXPERIMENTAL SEQUENCE}
\label{ExperimentalSequence}
Fig. 6 shows the experimental sequence that we use to load atoms into the ground band of the stroboscopic lattice.

\begin{enumerate}[label=\Roman*.]
	\item We start with atoms optically pumped into $|g_1\rangle$. We then ramp on $\Omega_{c1}$ (red trace in Fig.6) followed by $\Omega_p$ (blue trace in Fig.6), transferring atoms into a spatially homogeneous dark state. Then, we turn on $\Omega_{c2}(t)$ (green trace in Fig.6) in 200$\mu$s (minimum number of Floquet cycles used during the ramp $\approx$ 40) to adiabatically load atoms into the ground band of the stroboscopic lattice. 
	\item We pulse the stroboscopic lattice for a variable number of Floquet cycles.

	\item We measure the average probability density of the atoms in the ground band of the stroboscopic lattice using the nanoresolution microscopy technique described in Ref.~\cite{Subhankar2019}.
\end{enumerate}
The phase $\phi(t)$ of the $\Omega_{c2}$ light field, which controls the position of the stroboscopic lattice, is only changed when the dark-state spin composition is spatially homogeneous~\cite{Subhankar2019b}. The experimental techniques used to generate the pulses is detailed in Ref.~\cite{Subhankar2019}. 
\subsection{PULSE SCHEME}
\label{PulseScheme}

The functional form of $\Omega_{c2}(t)$ that we use to create the stroboscopic lattice is~\cite{Subhankar2019b}:
$$
\Omega_{c2}(t)=\frac{\Omega_{c0}}{2}-\frac{\Omega_{p}\sin^2(\omega_F t)}{\sqrt{1+4\epsilon^2-\sin^4(\omega_F t)}},$$
$$
\omega_F=\Omega_pr_0\sqrt{1+4\epsilon^2},
$$
where $\epsilon=\Omega_p/\Omega_{c0}$. In Fig.~4 , changes in $\omega_F$ are parameterized using $r_0.$ Smaller $r_0$ implies slower, more spin-adiabatic pulses. In our experiment, we typically use $0.02\le r_0\le0.2$. One Floquet period of pulsing is shown in Fig.~7. 
\begin{figure}[]
	\centering
	\includegraphics[width=8.6cm]{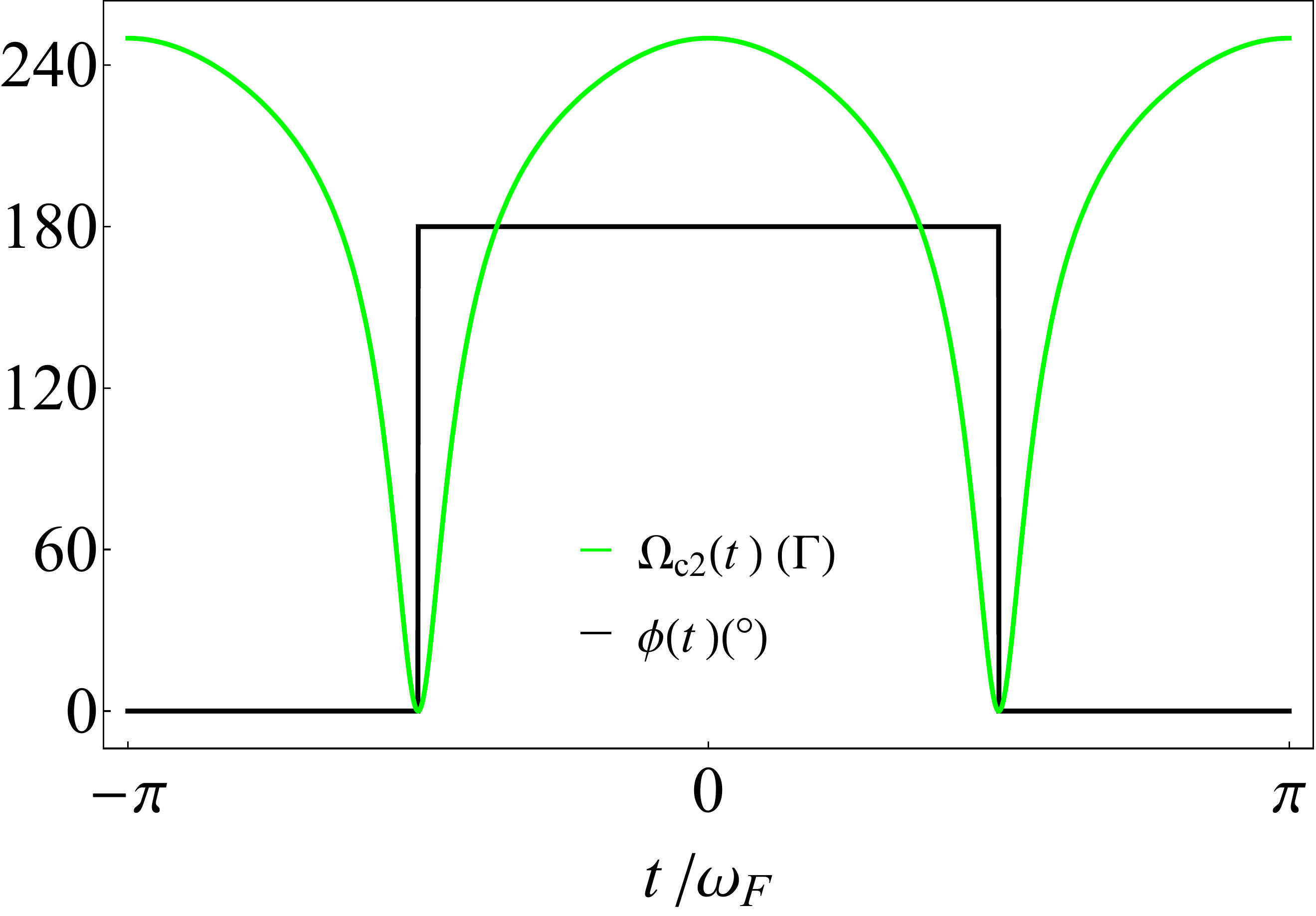}
	\label{fig:pulse} 
	\caption{One Floquet cycle of pulsing: The pulse shapes for $\Omega_{c2}(t)$ and $\phi(t)$. }
\end{figure}
\subsection{DETAIL OF LIFETIME STUDY}
\label{DetailofLifetimeStudy}
When studying lifetime for the STIRAP-only case and for the stroboscopic lattice case, we observe that $\sim20\%$ of the atoms have a lifetime of $\sim$ 20 ms and are insensitive to change in $\omega_F$. We speculate that these atoms populate Floquet states that are immune to STIRAP due to the large dynamic two-photon detunings arising from the spatially-dependent ac-Stark shifts due to couplings to states outside the $\Lambda$-system (Appendix~A). The decay rates shown in the main text pertain to the major fraction of the atoms which show frequency-dependent loss rates both in the stroboscopic lattice as well as stroboscopic STIRAP case.

\bibliography{apssamp}

\end{document}